\documentclass[
 reprint,
superscriptaddress,
 amsmath,amssymb,
 aps,
pra,
aip,
rsi,]{revtex4-1}
\usepackage{color}
\usepackage{graphicx}
\usepackage{caption}
\usepackage{subcaption}
\usepackage{dcolumn}
\usepackage{bm}
\usepackage{braket}
\usepackage{float}
\usepackage{multirow}
\usepackage{hhline}
\usepackage{mathtools}
\usepackage{qcircuit}
\usepackage{lipsum}

\begin{document}


\title{Optimizing Stabilizer Parities for Improved Logical Qubit Memories}
\author{Dripto M. Debroy}
\email{dripto@phy.duke.edu\\
Present Address: Google Research, Venice, CA 90291, USA}
\affiliation{Department of Physics, Duke University, Durham, NC 27708, USA}

\author{Laird Egan}
\affiliation{Joint Quantum Institute, Center for Quantum Information and Computer Science, and Departments of Physics and Electrical and Computer Engineering, University of Maryland, College Park, MD 20742, USA}
\email{Present Address: IonQ, College Park, MD 20740, USA}

\author{Crystal Noel}
\affiliation{Department of Physics, Duke University, Durham, NC 27708, USA}
\affiliation{Joint Quantum Institute, Center for Quantum Information and Computer Science, and Departments of Physics and Electrical and Computer Engineering, University of Maryland, College Park, MD 20742, USA}
\affiliation{Department of Electrical and Computer Engineering, Duke University, Durham, NC 27708, USA}

\author{Andrew Risinger}
\affiliation{Joint Quantum Institute, Center for Quantum Information and Computer Science, and Departments of Physics and Electrical and Computer Engineering, University of Maryland, College Park, MD 20742, USA}

\author{Daiwei Zhu}
\affiliation{Joint Quantum Institute, Center for Quantum Information and Computer Science, and Departments of Physics and Electrical and Computer Engineering, University of Maryland, College Park, MD 20742, USA}

\author{Debopriyo Biswas}
\affiliation{Joint Quantum Institute, Center for Quantum Information and Computer Science, and Departments of Physics and Electrical and Computer Engineering, University of Maryland, College Park, MD 20742, USA}

\author{Marko Cetina}
\affiliation{Department of Physics, Duke University, Durham, NC 27708, USA}
\affiliation{Joint Quantum Institute, Center for Quantum Information and Computer Science, and Departments of Physics and Electrical and Computer Engineering, University of Maryland, College Park, MD 20742, USA}

\author{Chris Monroe}
\affiliation{Department of Physics, Duke University, Durham, NC 27708, USA}
\affiliation{Joint Quantum Institute, Center for Quantum Information and Computer Science, and Departments of Physics and Electrical and Computer Engineering, University of Maryland, College Park, MD 20742, USA}
\affiliation{Department of Electrical and Computer Engineering,  Duke University, Durham, NC 27708, USA}
\affiliation{IonQ, College Park, MD 20740, USA}

\author{Kenneth R. Brown}
\email{ken.brown@duke.edu}
\affiliation{Department of Physics, Duke University, Durham, NC 27708, USA}
\affiliation{Department of Electrical and Computer Engineering,  Duke University, Durham, NC 27708, USA}

\begin{abstract}
We study variants of Shor's code that are adept at handling single-axis correlated idling errors, which are commonly observed in many quantum systems. By using the repetition code structure of the Shor's code basis states, we calculate the logical channel applied to the encoded information when subjected to coherent and correlated single qubit idling errors, followed by stabilizer measurement. Changing the signs of the stabilizer generators allows us to change how the coherent errors interfere, leading to a quantum error correcting code which performs as well as a classical repetition code of equivalent distance against these errors. We demonstrate a factor of $4$ improvement of the logical memory in a distance-3 logical qubit implemented on a trapped-ion quantum computer. Even-distance versions of our Shor code variants are decoherence-free subspaces and fully robust to identical and independent coherent idling noise. 
\end{abstract}

\pacs{Valid PACS appear here}
\maketitle

In quantum error correction, coherent errors are unwanted unitary operations applied to the physical qubits. Unlike stochastic errors, which scale linearly, coherent errors build up quadratically~\cite{kueng2016comparing,greenbaum2017modeling}. Coherent errors of various types can be mitigated through composite pulse sequences~\cite{viola1999dynamical,jones2003robust,brown2004arbitrarily}, random compiling~\cite{wallman2016noise, campbell2019random}, and circuit compilation~\cite{debroy2018slicing, cai2020mitigating}.

In this Letter, we consider errors resulting from spatially or temporally correlated phase noise. Such noise can arise from magnetic field fluctuations or instabilities in timing systems, which are a concern in most architectures, including trapped ions, superconductors, neutral atoms, and nitrogen-vacancy diamonds~\cite{ball2016role}. In optically addressed qubit systems, this type of noise can also appear due to beam path length fluctuations or finite laser linewidth~\cite{PhysRevA.100.062307}. Although quantum error correction will suppress these errors~\cite{beale2018coherence,huang2018performance}, they can increase logical qubit error relative to stochastic errors.

Previous work on temporally correlated idling error, also called coherent idling error, has focused on finding thresholds below which the coherence of the resulting logical channel is reduced~\cite{bravyi2017correcting, iverson2020coherence}. In Ref.~\cite{iverson2020coherence}, the authors present an exact solution for the logical channel experienced by a classical repetition code under a coherent idling error model. We use this simple model to solve for the exact logical channels of a select group of quantum error-correcting codes, which are variants of Shor's 9-qubit code~\cite{Shor1995}. Standard stabilizer codes stabilize even parity states. By changing the parity that a stabilizer preserves, we can directly control how the coherent errors constructively or destructively interfere. In this way, we can create codes where significant fractions of the coherent errors cancel out, similar to non-stabilizer code constructions~\cite{ouyang2020avoiding}. Even-distanced versions of our coherent error resilient code are members of the code family described in Ref. \cite{hu2020mitigating}, and have the errors fully cancel out. These codes are therefore robust to homogeneous coherent idling errors. The codespaces of these Shor-codes exist inside of a decoherence-free subspace~\cite{lidar1998decoherence,KnillPRL2000, kielpinski2001decoherence, deng2007preparation, andrews2019quantifying}.

Our calculations follow the example presented in Ref. \cite{iverson2020coherence}. Consider an error model where all qubits are rotated along the $Z$ axis by an angle $\theta$, represented by the channel $\mathcal{N}_\theta(\rho)$ on an $n$ qubit density matrix $\rho$:
\begin{equation}
    \begin{split}
        \mathcal{N}_\theta(\rho) &= Z(\theta)^{\otimes n}\rho (Z(\theta)^\dagger)^{\otimes n},\\
        Z(\theta) &= e^{-i\theta Z/2},
    \end{split}
    \label{eq:physerror}
\end{equation}
Now assume that the qubits are being used to encode one classical bit of information in a rotated $n$-bit repetition code:
\begin{equation}
    \begin{split}
        |+\rangle_L &= |+\rangle^{\otimes n},\\
        |-\rangle_L &= |-\rangle^{\otimes n}.\\
    \end{split}
    \label{eq:repcode}
\end{equation}
\begin{figure*}[ht] 
    \centering
    \includegraphics[width = 1.\linewidth]{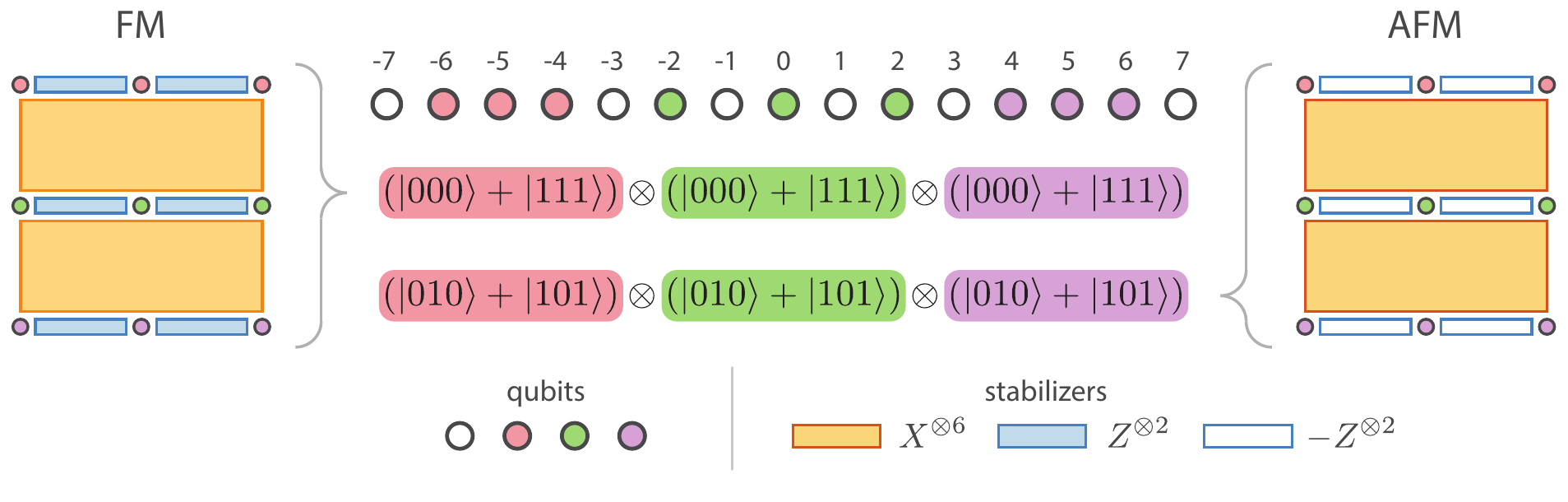}
    \caption{Diagrams for the ferromagnetic (left) and anti-ferromagnetic (right) [[$9$,$1$,$3$]] Shor's codes, as well as an ion chain with labeled ion indices. The qubit state representing $|0\rangle_L$ is shown for each variant, with each GHZ state color coded to match the ions in the chain as well as their location in the code diagrams. The qubits in each GHZ state are ordered from left to right.}
    \label{fig:ion mapping}
\end{figure*}

Once the error in Eq.~\ref{eq:physerror} has been applied, a round of stabilizer measurements are made, where the stabilizer generators of the repetition code are taken to be:
$$S_{rep} = \langle X_0X_1, X_1X_2,\ldots X_{n-2}X_{n-1}\rangle.$$
Every syndrome corresponds to two $Z$-type errors, related by $E_1 = E_2 Z_L$, where $Z_L = Z^{\otimes n}$. The correction applied is chosen by selecting the lower-weight error, which is the optimal decoding method if $\sin^2(\theta/2) < 1/2$. If we define $\alpha_s$($\beta_s$) as the prefactor to the correctable(uncorrectable) Pauli error corresponding to a syndrome $s$ in an expansion of Eq.~\ref{eq:physerror}, the logical channel after correction is:
\begin{equation}
    \begin{split}
        \mathcal{N}_{L}(\rho) &= \displaystyle\sum_s (\alpha_s I_L + \beta_s Z_L)\rho (\alpha_s^* I + \beta_s^* Z_L),\\
        &= \displaystyle\sum_s P_s \bar{Z}(\theta_s) \rho \bar{Z}(\theta_s)^\dagger.\\
    \end{split}
    \label{eq:replogicalchannel}
\end{equation}
\begin{equation}
    \begin{split}
        \bar{Z}(\theta) &\equiv  e^{-i\theta \bar{Z}/2},\\
        P_s &\equiv |\alpha_s|^2 + |\beta_s|^2,\\
        \theta_s &\equiv 2\arctan\left(\frac{i\beta_s}{\alpha_s}\right).
\end{split}
\end{equation}
As an example of this structure, for a 3-bit repetition code, the syndrome outcome of $01$ could be caused by an error $IIZ$ or an error $ZZI$. The weight-$1$ error is corrected, leading to
\begin{equation}
    \begin{split}
        \alpha_{01} &= \cos\left(\theta/2\right)^2\left(-i\sin\left(\theta/2\right)\right),\\
        \beta_{01} &= \cos(\theta/2)(-i\sin(\theta/2))^2.
    \end{split}
\end{equation}
These amplitudes lead to a rotation angle of
\begin{equation}
    \theta_{01} = 2\arctan\left( \frac{\sin(\theta/2)}{\cos(\theta/2)}\right) = \theta,
\end{equation}
meaning that for this syndrome the logical rotation angle is the same as the physical rotation angle.

This is not always true, and as shown in Eq.~\ref{eq:replogicalchannel}, the logical channel is a logical $Z$ rotation with an angle conditional on the syndrome outcome measured. In the case of an $n$-bit repetition code, the values of $\alpha_s$ and $\beta_s$ only depend on $n$ and the weights of the corresponding errors, and are completely independent of the error arrangement. As a result, one can define the quantities:
\begin{equation}
    \begin{split}
        P_{n,w}(\theta) &= {n \choose w}((\cos(\theta/2)^{(n-w)}\sin(\theta/2)^{w})^2\\
        &\hspace{8ex}+ (\cos(\theta/2)^{w}\sin(\theta/2)^{(n-w)})^2),\\ \\
        \theta_{n,w} &= (-1)^{(n-2w-1)/2}2\arctan(\tan^{n-2w}(\theta/2)),
    \end{split}
    \label{eq:p_nw theta_nw}
\end{equation}
where $n$ is the distance of the repetition code and $w$ is the weight of the correctable (lower weight) error. The logical channel in Eq.~\ref{eq:replogicalchannel} can then be rewritten as:
\begin{equation}
    \mathcal{N}_{L}(\theta) = \displaystyle\sum_{w = 0}^{(n-1)/2} P_{n,w}(\theta) U_{\bar{Z}}(\theta_{n,w})\rho U_{\bar{Z}}(\theta_{n,w})^\dagger.
    \label{eq:nw rep code log channel}
\end{equation}

This compact description of the logical channel seen by the repetition code relies on the simple construction of the code. For most quantum error correcting codes, syndromes do not translate as directly into easily understood errors. We study the case of Shor's codes, which do follow this structure. The 9-qubit code presented in  Ref.~\cite{Shor1995} can be thought of as three 3-bit repetition codes with $Z$-type stabilizers, concatenated into a repetition code with $X$-type stabilizers. The resulting code, with 6 weight-2 $Z$-type stabilizers and 2 weight-6 $X$-type stabilizers, has a structured ground state composed of a product of Greenberger-Horne-Zeilinger (GHZ) states:
\begin{equation}\label{eq:shor states}
    \begin{split}
        |0\rangle_L &\equiv \frac{1}{2\sqrt{2}}\left(|000\rangle + |111\rangle\right)^{\otimes 3} ,\\
        |1\rangle_L &\equiv \frac{1}{2\sqrt{2}}\left(|000\rangle - |111\rangle\right)^{\otimes 3}.\\
    \end{split}
\end{equation}
In this paper we discuss this code, as well as a variant with $Z$-type stabilizer generators of opposite parity. In Appendix~\ref{app:shorbd} we discuss the pair of codes created by taking the two codes presented here and swapping the stabilizer bases. 

On the left of Fig.~\ref{fig:ion mapping} is the standard 9-qubit Shor's code. The logical state preparation and measurement of this code has been demonstrated with trapped ions \cite{egan2020fault,nguyen2021demonstration} and photons \cite{luo2020quantum}.  The phase errors on a given GHZ state combine constructively, as can be seen for the states shown in Eq.~\ref{eq:shor states}. This is a consequence of the $ZZ$ stabilizers along each row leading to $Z(\theta)$ errors being indistinguishable for qubits on the same row. We can imagine pushing all the errors to the leftmost column of qubits. Each of these qubits will experience a rotation with an angle of $n\theta$, and the outer repetition code will have the same logical channel as an $n$-bit logical channel in Eq.~\ref{eq:nw rep code log channel}, with $\theta \rightarrow n\theta$:
\begin{equation}
    \mathcal{N}_{L}(\theta) = \displaystyle\sum_{w = 0}^{(n-1)/2} P_{n,w}(n\theta) U_{\bar{Z}}(n\theta_{n,w})\rho U_{\bar{Z}}(n\theta_{n,w})^\dagger.
    \label{eq:nw ntheta shor code log channel}
\end{equation}
This represents a worst-case situation where the logical channel will experience an error that increases quadratically with distance.

We now consider the code on the right of Fig.~\ref{fig:ion mapping}, which we will refer to as the anti-ferromagnetic case since the GHZ states composing its logical states resembles the ground states of an anti-ferromagnetic Ising spin chain. This code has the same stabilizer structure as that on the left of Fig.~\ref{fig:ion mapping}, but with the signs of the $ZZ$ stabilizers changed. This will not have any impact on the code's ability to correct stochastic errors, however it causes the interference of the coherent idling errors to go from constructive to destructive. In the even-distance case, all the errors cancel out. These codes are consequently immune to error of the type described in Eq.~\ref{eq:physerror}, and are an example of the codes described in Ref.~\cite{hu2020mitigating}. For the odd-distanced cases, a single error does not cancel on each row. As a result, the effective error channel seen by the outer repetition code is identical to that seen in the classical repetition code case, so the logical error channel is identical to that seen in Eq.~\ref{eq:nw rep code log channel}. This represents a significant improvement in the logical error channel, as the repetition code has the maximum possible threshold of $p_{th} = 1/2$.

Our discussion above assumes temporal and spatial correlation, but these modified codes also improve protection for errors that are only spatially correlated. This follows from previous work on correlated dephasing noise in the context of weak decoherenece free subspaces~\cite{KempePRA2001}.  Here we do not seek perfect cancellation of correlated errors, but instead a linear reduction in the error rate.

To confirm these results, we test them on a trapped ion quantum computer that has previously demonstrated fault-tolerant error-correction protocols~\cite{egan2020fault}. A chain of 15 ions is trapped above a microfabricated chip trap~\cite{Maunz2016}, with optical individual site addressing controlled by a multi-channel acousto-optic modulator. Measurement, single-qubit gate, and two-qubit gate fidelities are $>99.5\%$, $99.98\%$, and $98.5-99.3\%$ respectively~\cite{egan2020fault}. The qubit in this system is defined on the electronic ground state hyperfine ``clock" states of $^{171}$Yb$^+$ ions: $|0\rangle \equiv |F=0;m_F=0\rangle$, $|1\rangle\equiv |F_1;m_F=0\rangle$. The qubit frequency is $\omega_0 = 2\pi \times 12,642,812,118.5 + \delta_2$~ Hz where $\delta_2= (310.8)B^2$~Hz is the second-order Zeeman shift for a magnetic field $B$ in Gauss~\cite{fisk1997accurate}. 

In our system, residual second-order sensitivity to magnetic fluctuations and/or local oscillator noise limits the $T_2$ decoherence time to $\approx2.75$~s, whereas the qubit itself is capable of $T_2>1$~hour~\cite{wang2020single}. The relevant quantity for this paper is the un-echoed $T_2^*=0.6$ s decoherence time using optical control of the qubits, likely dominated by mechanical vibrations that shift the phase of the optical standing wave relative to the ion. In contrast to phase noise, there is also a well-characterized and static linear magnetic field gradient across the length of ion chain that results in a qubit frequency shift of $\pm 4$~Hz shift relative to the center ion. In Ref. \cite{egan2020fault} the chip was rotated so that the magnetic field was constant throughout the chip, as opposed to the linearly varying magnetic field in this geometry. At the circuit level, if this shift is not accounted for in software then it will create relative phase shifts between the ions that may also appear as a coherent idling error. 

We first study the individual ferromagnetic and anti-ferromagnetic GHZ states which compose the logical states. These states are:
\begin{equation}
    \begin{split}
        |FM_{n}\rangle &= \frac{1}{\sqrt{2}}\left(|000\ldots\rangle + |111\ldots\rangle\right),\\
        |AFM_{n}\rangle &= \frac{1}{\sqrt{2}}\left(|010\ldots\rangle + |101\ldots\rangle\right),
    \end{split}
\end{equation}
where it is now clear to see the resemblance between the FM and AFM states and the ground states of the ferromagnetic and anti-ferromagnetic spin chains, respectively. A code with $n\times n$ qubits would have a logical state which is a tensor product of $n$ of these states. The logical states and ion-to-qubit mappings for the [[$9$,$1$,$3$]] ferromagnetic and anti-ferromagnetic Shor's codes are shown in Fig.~\ref{fig:ion mapping}. 

\begin{figure}
    \centering
    \includegraphics[width = \linewidth]{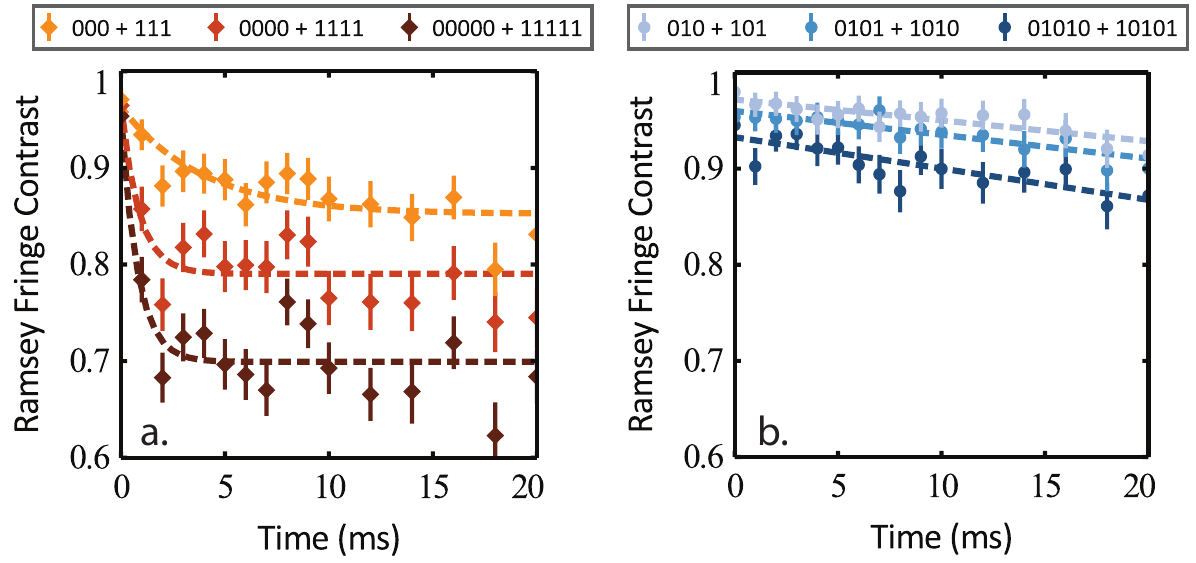}
    \caption{Fringe contrast after a Ramsey experiment on the (a) ferromagnetic and (b) anti-ferromagnetic GHZ states for different distances. The dashed lines are present to serve as guides to the eye.}
    \label{fig:exp - ghz}
\end{figure}

After these states are prepared, we perform a Ramsey experiment with variable wait time to measure the coherence of the state as a function of time, as in Ref. \cite{egan2020fault}. The result of this experiment for different GHZ sizes is shown in Fig.~\ref{fig:exp - ghz}. For the ferromagnetic states in Fig.~\ref{fig:exp - ghz}a, we observe an initial fast decay of the contrast on a timescale that corresponds to the correlation time of phase noise in our system, followed by a longer slow decay. The anti-ferromagnetic states in Fig.~\ref{fig:exp - ghz}b, which correspond to codes like the one shown on the right of Fig.~\ref{fig:ion mapping}, exhibit only a slow decay, which indicates that these states are more resistant to the correlated phase errors present in the system. Using these states allows the codes designed on top of them to inherit their robustness. While the $4$-qubit AFM GHZ state is predicted to be completely insensitive to our dominant idling error, we do observe a small decay over 10~ms, indicating that there may be small contributions from a different error source. As shown, the increased robustness to coherent idling error relative to the 3-qubit AFM GHZ state is not quite enough to outweigh the cost of an additional qubit and entangling gate. 
\begin{figure}
    \centering
    \includegraphics[width = \linewidth]{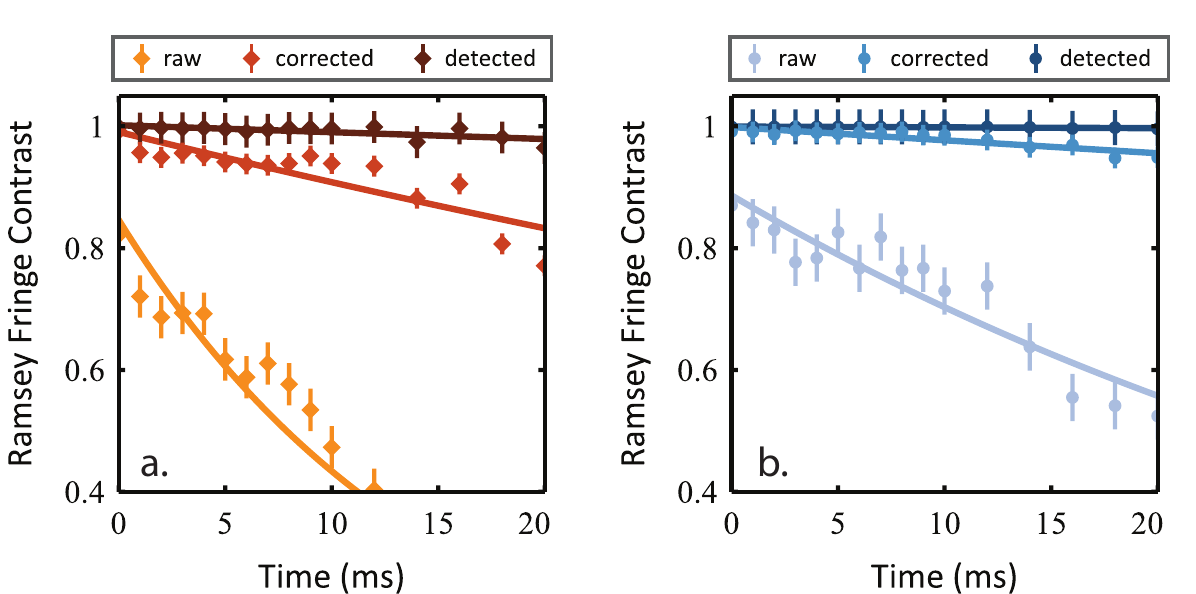}
    \caption{Experimental dephasing performance of (a) ferromagnetic and (b) anti-ferromagnetic [[$9$, $1$, $3$]] Shor's code logical states. The Ramsey fringe amplitude gives the coherence of the states. The data is fit to $A\exp(-\Gamma t)$. Fitting parameters are presented in Appendix~\ref{app:fits}.}
    \label{fig:exp - bs13}
\end{figure}

We can now directly compare the performance of FM and AFM versions of a full [[$9$,$1$,$3$]] Shor's code in Fig.~\ref{fig:exp - bs13}, using a logical qubit Ramsey experiment identical to the one performed in Ref.~\cite{egan2020fault}. To take this data, three separate 3-qubit GHZ states are constructed simultaneously, as shown in Fig.~\ref{fig:ion mapping}. Using the structure of the codestates, we apply one round of error correction based off the measurement outcomes of the data qubits~\cite{egan2020fault}, leading to the three curves in Fig.~\ref{fig:exp - bs13}. Raw curves are constructed by the total parity of the 9 data qubits after a measurement in the $X$ basis, error corrected curves reconstruct the stabilizer outcomes from these measurements and apply one correction, while error detected curves reconstruct these stabilizer outcomes and then discard any run in which the stabilizers are violated. In this manner, error correction corresponds to ``post-processing" of the data, whereas error detection corresponds to ``post-selection" of the data. A single error on the data qubits flips the outcome of the raw data, two errors are required to flip the outcome of the corrected data, and three errors are needed to flip the outcome of the detected data. We can see that the ferromagnetic code performs worse than the anti-ferromagnetic code in all three cases. From this data we can calculate logical $T2^*$ times of $115(10)$~ms in the ferromagnetic case and $450(150)$~ms in the anti-ferromagnetic case.
\begin{figure}[t!]
    \centering
    \includegraphics[width =\linewidth]{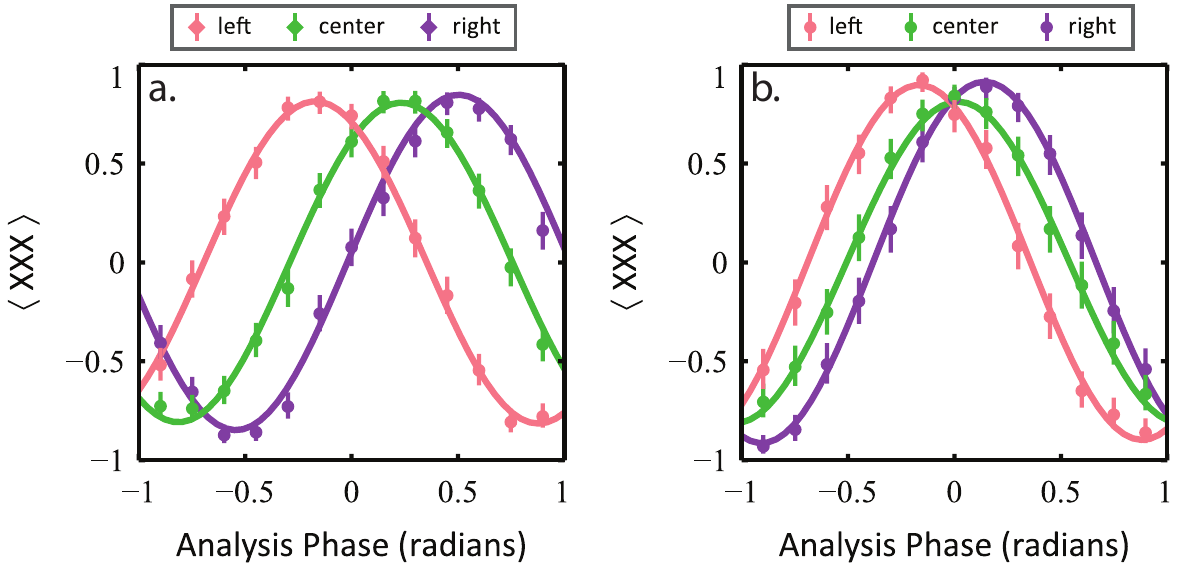}
    \caption{Individual GHZ state Ramsey fringes after a $20$~ms wait time for the (a) ferromagnetic and (b) anti-ferromagnetic [[$9$,$1$,$3$]] Shor's code $|0\rangle_L$ logical states. Colors are selected to match the GHZ state coloring from Fig.~\ref{fig:ion mapping}. The data is fit to $A\cos(3\phi + \phi_0)$. Fitting parameters are presented in Appendix~\ref{app:fits}.}
    \label{fig:exp - 3ghz}
\end{figure}    

The logical states presented in Eq.~\ref{eq:shor states} are composed of three separate GHZ states, which allows us to study their performance separately as presented in Fig.~\ref{fig:exp - 3ghz}. We note that the reduction in contrast for the central AFM GHZ state is due to lower gate fidelities in its preparation circuit, not coherent error. By considering the spatial arrangement of the GHZ states, as shown in Fig.~\ref{fig:ion mapping}, we can study the impact of the static magnetic field gradient on the code. It should be noted that any state which balances the number of excitations in each GHZ state is resilient to dynamic noise of the form described in Eq.~\ref{eq:physerror} and shown in Fig. \ref{fig:exp - ghz}. Our particular mapping, however, is also robust to magnetic fields which slowly vary in space because our states have errors cancel with their nearest neighbors. In Fig.~\ref{fig:exp - 3ghz}, we see that the magnetic field appears to vary linearly across the axis of the trap, leading to GHZ states experiencing different phase shifts depending on their position in the chain. While this error does not decrease the coherence of individual GHZ states, it does affect the performance of the code, which depends on the three GHZ states remaining in phase with each other. The physical states shown in Fig.~\ref{fig:exp - 3ghz}a, which correspond to the FM code in Fig.~\ref{fig:ion mapping}, experience higher differential phase shifts than their counterparts from Fig.~\ref{fig:exp - 3ghz}b, which correspond to the AFM code. If we assume that the magnetic field is linearly increasing along the chain, we can define the angle of a qubit at integer position $x$ as
\begin{equation}
    \theta_x = \theta_0 + x\delta,
\end{equation}
where $\delta$ is the difference in phase between a qubit in position $x$ and the one in $x+1$, and $\theta_0$ is the phase accumulated by the central ion. Under these conditions, the phase accumulated for the FM GHZ states in Fig.~\ref{fig:ion mapping} would be:
\begin{equation}
    \begin{split}
        \theta_{\mbox{left}} &= \theta_{-6} + \theta_{-5} + \theta_{-4} = 3\theta_0 - 15\delta,\\
        \theta_{\mbox{center}} &= \theta_{-2} + \theta_{0} + \theta_{2} = 3\theta_0,\\
        \theta_{\mbox{right}} &= \theta_{4} + \theta_{5} + \theta_{6} = 3\theta_0 + 15\delta.
    \end{split}
\end{equation}
In comparison, the phase accumulated for the AFM GHZ states would be:
\begin{equation}
    \begin{split}
        \theta_{\mbox{left}} &= \theta_{-6} - \theta_{-5} + \theta_{-4} = \theta_0 - 5\delta,\\
        \theta_{\mbox{center}} &= \theta_{-2} - \theta_{0} + \theta_{2} = \theta_0,\\
        \theta_{\mbox{right}} &= \theta_{4} - \theta_{5} + \theta_{6} = \theta_0 + 5\delta.
    \end{split}
\end{equation}
This corresponds to a threefold reduction in accumulated phase for the individual GHZ states. These accumulated phases take the place of $\theta$ in Eq.~\ref{eq:replogicalchannel} when error correction is applied using a distance-3 code, a threefold reduction in accumulated phase would correspond to an approximately $81$-fold reduction in logical error rate, assuming no other error sources existed in the system.

In Appendix~\ref{app:0, -2, 2} we discuss a variation on this experiment which allowed us to experimentally confirm our understanding of the magnetic fields present in our system. Of course, if the magnetic fields are static and well known, a preferable option would be to adjust the qubit frequencies in classical control. However this option is not possible for unknown drifts which may occur during a computation. Additionally, in trapped-ion architectures that involve extensive shuttling operations~\cite{pino2020demonstration}, qubits will often acquire a path dependant phase that needs to be calibrated and pre-calculated. Codes that are robust to this error may reduce calibration overheads and circuit compilation complexity.

Our primary conclusion is that changing stabilizer parity allow us to control the interference between correlated idling errors. We present a family of codes, which we refer to as anti-ferromagnetic Shor's codes, which inherit the one sided threshold of standard Shor's codes while also possessing a threshold against correlated idling noise in the other basis. The even-distanced versions of our code family are an example of the larger set of codes described in Ref.~\cite{hu2020mitigating}. The particular choice of qubit mapping we use also resistant to magnetic fields which are slowly varying in space, like the ones seen in many systems. Such idling-resistant codes could be used in a concatenated scheme much like the one discussed in Ref.~\cite{ouyang2020avoiding}. We present experimental data from a trapped ion quantum computer which demonstrates our codes showing marked improvements in performance, relative to the standard Shor's code. We can also change parity preserved by higher weight stabilizers, as considered in Appendix~\ref{app:shorbd}, leading to less drastic cancellations. These modifications would still leave error correcting performance against uncorrelated stochastic errors unchanged, while improving the resilience to correlated idling error, and could be implemented on top of any CSS code.
\vspace{1ex}
\section{Acknowledgements}
This work was performed at the University of Maryland (UMD), with no laboratory or equipment support from IonQ. The authors thank Michael Newman, Robert Calderbank, Jingzhen Hu, Qingzhong Liang, and Narayanan Rengaswamy for helpful conversations. This work was supported by the Office of the Director of National Intelligence - Intelligence Advanced Research Projects Activity through ARO contract W911NF-16-1-0082, National Science Foundation Expeditions in Computing award 1730104, National Science Foundation STAQ project Phy-1818914, and ARO MURI grant W911NF-18-1-0218. D.M.D. and L.N.E. are also funded in part by NSF QISE-NET fellowships (DMR-1747426). 

\bibliographystyle{apsrev}
\bibliography{References}
\clearpage
\onecolumngrid
\appendix
\section{Changing the Sign of  Higher Weight Stabilizers}\label{app:shorbd}
\begin{figure}[ht]
    \centering
    \includegraphics[width = 0.5\linewidth]{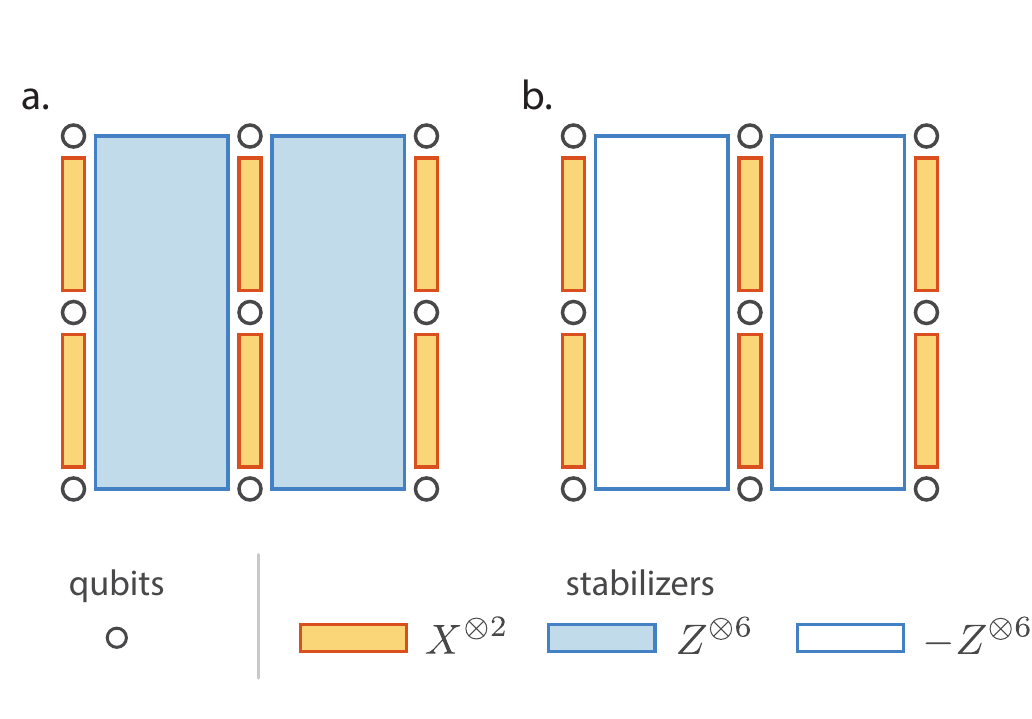}
    \caption{The two variants of the [[$9$,$1$,$3$]] Shor's codes presented in Fig.~\ref{fig:ion mapping}, with reversed stabilizer bases. Orange stabilizers are $\prod_i X_i$ stabilizers, filled blue stabilizers are $\prod_iZ_i$ stabilizers, and empty blue stabilizers are $-\prod_iZ_i$ stabilizers. In (a) we show a Shor's code with weight-6 $Z$ stabilizers that are all positive, while in (b) we present a variant of the Shor's code shown in (a), with the $Z$ stabilizers negated to improve performance against coherent idling errors.}
    \label{fig:shorbd}
\end{figure}

In Fig.~\ref{fig:shorbd}a, we present a variant of the original Shor's code with the $X$ and $Z$ stabilizers interchanged, making it more optimized to detect $Z$-type errors. This has the side effect of also preventing the coherent idling errors from interfering constructively. Since any odd number of the individual GHZ states experiencing a $Z$-type logical error would lead to a logical error on the full code, the logical channel of the total code corresponds to the same logical channel from Eq.~\ref{eq:nw rep code log channel} being applied $n$ times, where $n$ is the distance of the code. The logical rotations would add coherently, so the total channel can be written as:
\begin{equation*}
    \mathcal{N}_{L, shor}(\theta) = \displaystyle\sum_{w_1,\ldots w_{n-1} = 0}^{(n-1)/2} P_{n,\overrightarrow{w}} U_{\bar{Z}}(\theta_{n,\overrightarrow{w}})\rho U_{\bar{Z}}(\theta_{n,\overrightarrow{w}})^\dagger,
\end{equation*}
\begin{equation}
    \begin{split}
        P_{n,\overrightarrow{w}} &= \displaystyle\prod_i P_{n, w_1},\\
        \theta_{n,\overrightarrow{w}} &= \displaystyle\sum_i \theta_{n, w_i}
    \end{split}
    \label{eq:shor's standard log channel}
\end{equation}
where the quantities $P_{n,w_i}$ and $\theta_{n,w_i}$ were defined in Eq.~\ref{eq:p_nw theta_nw}. Each of the individual GHZ states making up the logical state experiences its own rotation, and the angles are summed up to find the total angle experienced by the code.

Next, we can consider the case presented in Fig.~\ref{fig:shorbd}b, which is a variant of Fig.~\ref{fig:shorbd}a with the signs of all the weight-$2n$ $Z$ stabilizer generators changed. This results in a logical channel like the one in Eq.~\ref{eq:shor's standard log channel}, but with the logical rotation angles having alternating signs:
\begin{equation}
    \displaystyle\sum_i \theta_{n, w_i} \rightarrow \displaystyle\sum_i (-1)^i\theta_{n, w_i}.
\end{equation}
The angles $\theta_{n,w}$ can be either positive or negative, but for finite distances the distribution is not perfectly symmetric. As a result, summed angles in the logical channel for the code in Fig.~\ref{fig:shorbd}a can be described by a biased random-walk, while the scrambling of signs in Fig.~\ref{fig:shorbd}b removes some of this bias. For even distances, the angle would be the difference of two equally biased random-walks, while for odd distances one side would have a single extra step. As a result, the changing the signs of the stabilizer generators significantly reduce the overall logical rotation angle by mitigating the bias.

This performance increase can also be understood as destructive interference of higher-order terms in Eq.~\ref{eq:physerror}. If the stabilizer $Z_0Z_1Z_3Z_4Z_6Z_7$ is changed to $-Z_0Z_1Z_3Z_4Z_6Z_7$, the term in Eq.~\ref{eq:physerror} corresponding to the error $Z_0Z_1Z_3$ would cancel with the term corresponding to $Z_4Z_6Z_7$. Pairs with unequal weights, like $Z_0Z_7$ and $Z_1Z_3Z_4Z_6$ would destructively interfere as well, but the unequal prefactors would result in some residual error.

\begin{figure}
    \centering
    \includegraphics[width=0.5\linewidth]{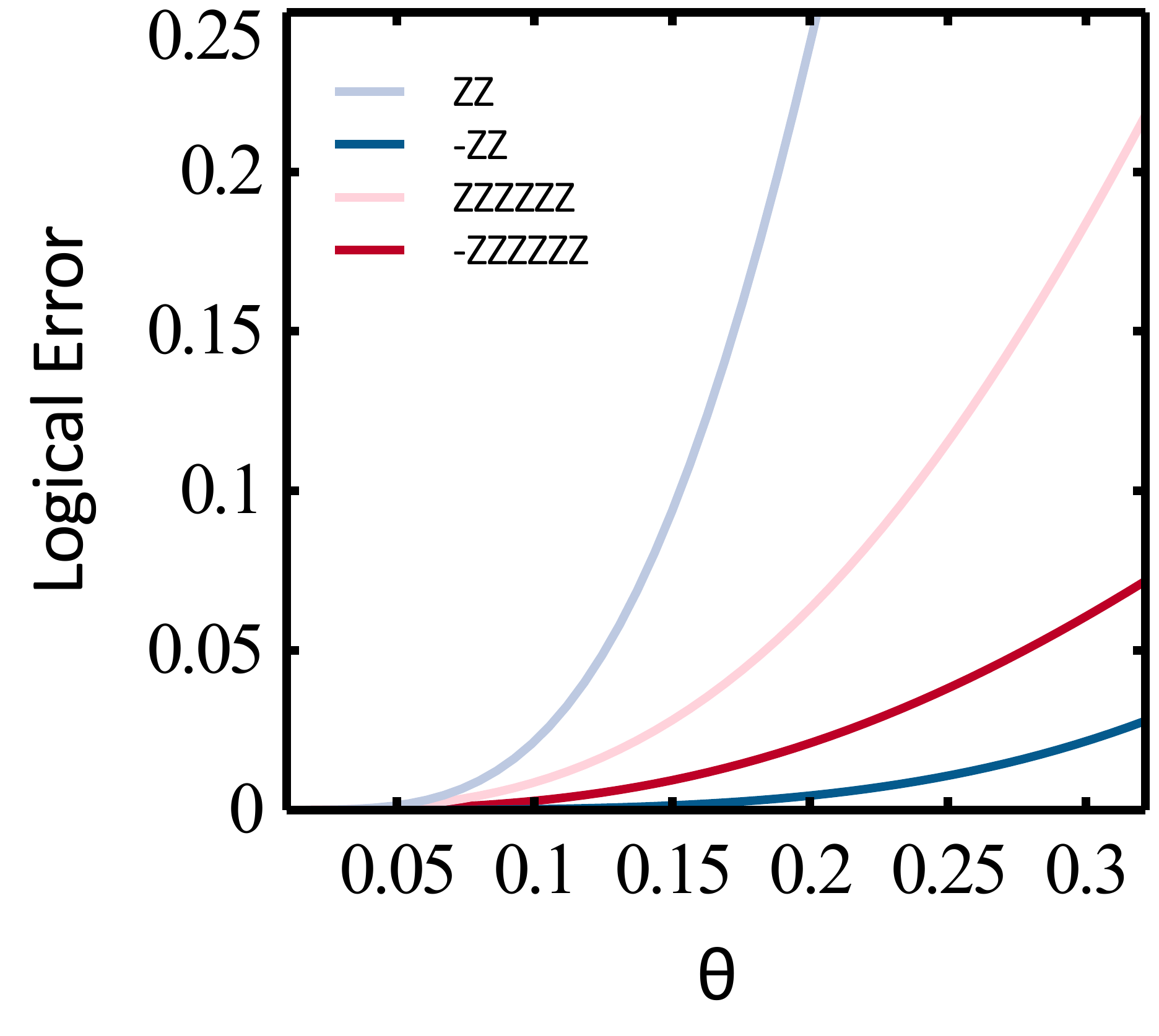}
    \caption{Logical error after a single round for the four codes presented in the main text and Appendix \ref{app:shorbd}, with codes labeled by their $Z$-type stabilizer generators. The codes in the main text (FM shown in light blue, AFM in dark blue), have direct interference between errors, leading to more extreme behavior than the ones presented in Appendix \ref{app:shorbd}, which only have interference between higher weight error terms.}
    \label{fig:logerrs}
\end{figure}

To understand how the two codes presented in the main text compare to the two codes presented in this Appendix, in Fig.~\ref{fig:logerrs} we plot their logical error after a single round. The codes with higher weight $Z$-type stabilizers show less improvement due to changing stabilizer signs, but still show an improvement.

\section{Experimental Fits}\label{app:fits}
\begin{table}[h]
    \centering
    \begin{tabular}{c|l|c|c}
        Figure & Dataset & $A$ & $\Gamma$\\ 
        \hline
        Fig.~\ref{fig:exp - bs13} & FM - raw & $0.846$ & $0.0668$ \\
        Fig.~\ref{fig:exp - bs13} & FM - corrected & $0.991$ & $0.00870$ \\
        Fig.~\ref{fig:exp - bs13} & FM - detected & $1.00$ & $0.00114$ \\
        Fig.~\ref{fig:exp - bs13} & AFM - raw & $0.887$ & $0.0232$ \\
        Fig.~\ref{fig:exp - bs13} & AFM - corrected & $0.999$ & $0.00220$ \\
        Fig.~\ref{fig:exp - bs13} & AFM - detected & $1.00$ & $0.000131$ \\
    \end{tabular}
    \caption{Fitting parameters for Fig.~\ref{fig:exp - bs13}. The curves are fit to $A\exp(-\Gamma t)$ with $t$ in milliseconds.}
    \label{tab:t2 fits}
\end{table}

\begin{table}[h]
    \centering
    \begin{tabular}{c|l|c|c}
        Figure & Dataset & $A$ & $\phi_0$\\
        \hline
        Figs.~\ref{fig:exp - 3ghz} \& \ref{fig:0, -2, 2} & FM - left & $0.814$ & $0.512$\\
        Figs.~\ref{fig:exp - 3ghz} \& \ref{fig:0, -2, 2} & FM - center & $0.808$ & $2.44$\\
        Figs.~\ref{fig:exp - 3ghz} \& \ref{fig:0, -2, 2} & FM - right & $0.848$ & $1.63$\\
        \hline
        Fig.~\ref{fig:exp - 3ghz} & AFM - left & $0.897$ & $0.498$\\
        Fig.~\ref{fig:exp - 3ghz} & AFM - center & $0.813$ & $3.13$\\
        Fig.~\ref{fig:exp - 3ghz} & AFM - right & $0.913$ & $2.72$\\
        \hline
        Fig.~\ref{fig:0, -2, 2} & AFM - left & $0.911$ & $3.13$\\
        Fig.~\ref{fig:0, -2, 2} & AFM - center & $0.890$ & $2.49$\\
        Fig.~\ref{fig:0, -2, 2} & AFM - right & $0.91$ & $2.54$\\
    \end{tabular}
    \caption{Fitting parameters for Figs.~\ref{fig:exp - 3ghz} and \ref{fig:0, -2, 2}. The curves are fit to $A\cos(3\phi + \phi_0)$.}
    \label{tab:cos fits}
\end{table}

In Table~\ref{tab:t2 fits} we present the fits for the data shown in Fig.~\ref{fig:exp - ghz} and Fig.~\ref{fig:exp - bs13}, which are all exponentially decaying fits of the form $A\exp(-\Gamma t)$. The amplitude, $A$, corresponds to the fidelity of the preparation circuits for the states we are interested in, while $\Gamma$ is the rate of decay in the system which corresponds to $\Gamma = 1 / T_2^*$.

In Table~\ref{tab:cos fits} we present the fits for the data in Fig.~\ref{fig:exp - 3ghz} and Fig.~\ref{fig:0, -2, 2}, which are fit to $A\cos(3\phi + \phi_0)$. Here, $A$ corresponds to the amplitude of the cosine curve, which tells us how much stochastic dephasing has occurred, as unitary dephasing would affect the phase through a shift of $\phi_0$.



\section{0, -2, 2 state data}\label{app:0, -2, 2}
\begin{figure}
    \centering
    \includegraphics[width = 0.7\linewidth]{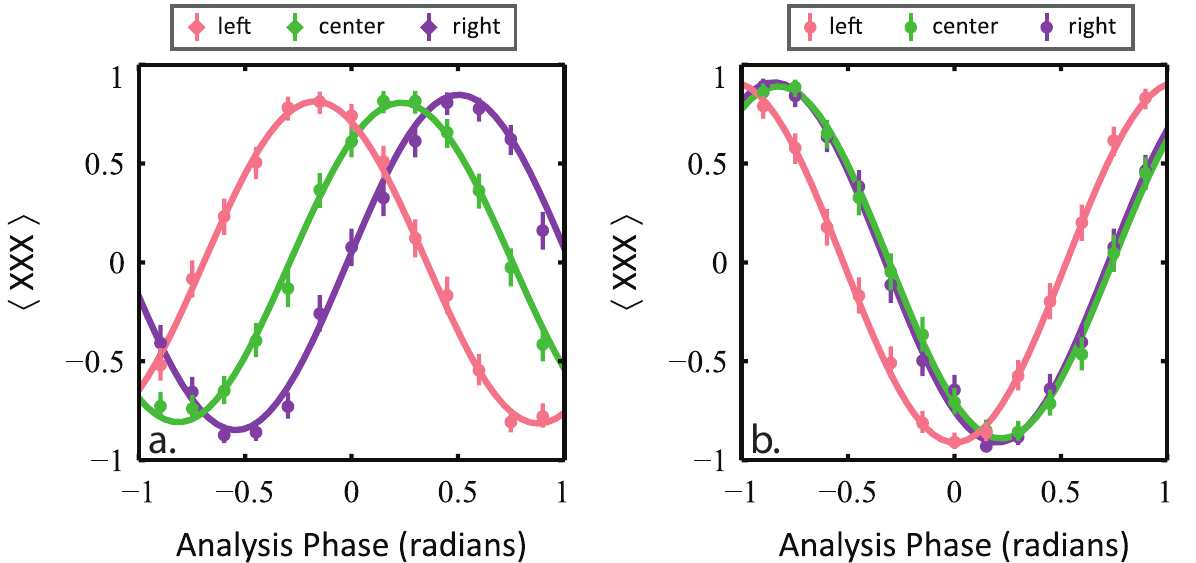}
    \caption{In (a) we replicate the FM GHZ state performance from Fig.~\ref{fig:exp - 3ghz}a. In (b) we present the performance of the anti-ferromagnetic [[$9$,$1$,$3$]] Shor's code $|0\rangle_L$ logical states under a modified ion-to-qubit mapping. Colors are selected to match the GHZ state coloring from Fig.~\ref{fig:ion mapping}, with red being the state on the qubits in positions $\{-6, -5, -4\}$, green being the state on the qubits in positions $\{0, -2, 2\}$, and purple being the state on the qubits in positions $\{4, 5, 6\}$. The data is fit to $A\cos(3\phi + \phi_0)$. Fitting parameters are presented in Appendix~\ref{app:fits}.}
    \label{fig:0, -2, 2}
\end{figure}
In Fig.~\ref{fig:0, -2, 2}, we present the GHZ data for the FM and AFM GHZ state idling performance for a slightly modified version of the results in the main text. The center GHZ state has been remapped to a qubit ordering of $\{0, -2, 2\}$. This leaves the FM GHZ state unaffected, while modifying the AFM GHZ state to $\frac{1}{\sqrt{2}}\left( |011\rangle + |100\rangle\right)$ for the central triplet of qubits, when ordered by their position in the chain.

Following the same analysis as the main text, we can calculate $\theta_{center}$ for the this modified variant of the AFM state:
\begin{equation}
    \theta_{center} = \theta_{0} - \theta_{-2} + \theta_{2} = \theta_0 + 4\delta,
\end{equation}
which is close to $\theta_{right} = \theta_0 + 5\delta$. Note that $\theta_{center}$ stays the same for the FM case.

We can see that this hypothesis matches nicely with the experimental results. In the FM case, the plot in Fig.~\ref{fig:0, -2, 2} looks the same as the FM case for the standard ordering in Fig.~\ref{fig:exp - 3ghz}. However in the AFM case, the center AFM state tracks closely to the right state, which had a dephasing angle of $\theta_{right} = \theta_0 + 5\delta$. This confirms our belief that the static magnetic field in our system is approximately linear in axial position, while also showing that the standard $\{-2, 0, 2\}$ mapping is preferable for creating codestates robust against this particular field. 
\clearpage

\end{document}